Whilst this Planet Has Gone Cycling On: What Role for Periodic Astronomical Phenomena in Large Scale Patterns in the History of Life?

By Bruce S. Lieberman[1] and Adrian L. Melott[2]
[1]Department of Geology, and Natural History Museum/Biodiversity Research Center,
1475 Jayhawk Blvd., 120 Lindley Hall
[2]Department of Physics and Astronomy, 1251 Wescoe Dr. #1082,
University of Kansas, Lawrence, KS, 66045, U.S.A.

**Abstract**

One of the longstanding debates in the history of paleontology focuses on the issue of whether or not there have been long term cycles (operating over 10's of millions of years) in biodiversity and extinction. Here we consider the history of this debate by connecting the skein from Grabau up to 2008. We focus on the evidence for periodicity that has emerged thus far, and conclude that there is indeed some evidence that periodicity may be real, though of course more work is needed. We also comment on possible causal mechanisms, focusing especially on the motion of our solar system in the Galaxy. Moreover, we consider the reasons why some scientists have opposed periodicity over the years. Finally, we consider the significance of this for our understanding of evolution and the history of life.

X.1 Introduction

Charles Darwin deserves a great deal of the credit for convincing people that there are natural processes continually and uniformly acting to shape biological evolution. One of the eloquent ways that he used to try to sway people was by invoking the analogy of how gravity continually acts to maintain the structure of the solar system: hence the genesis of the title of this paper, which incorporates a fragment from the last sentence of his epochal "On the Origin of Species … "(Darwin 1859, p. 490). Of course, scientists have long recognized the clockwork nature of the nearby confines of our solar system; further, since the middle of the 20$^{th}$ century, the bulk of the biological community has accepted the primary role of natural selection as the shaper of evolution. Thus, Darwin's analogy between astronomy and biology has been universally accepted; however, documenting a causal link between cyclical astronomical phenomena and periodic evolution and extinction has proven far more tenuous and controversial. Here we try to forge this causal link and consider the question of how large scale, cyclical astrophysical processes may have influenced the history of life at the grand scale, defined here (perhaps arbitrarily) as over the course of tens of millions of years. Such a thesis no longer seems as farfetched as it once did, now that paleontological studies have shown that astronomical objects, like bolides (at the end of the Cretaceous, Alvarez et al. 1980), and astrophysical phenomena, like gamma ray bursts (at the end of the Ordovician, Melott et al. 2004), may have precipitated mass extinctions (though not necessarily repeatedly and cyclically). Events like bolide impacts at the end of the Cretaceous powerfully influence the history of life, in the manner Gould (1989) referred to as *contingency* because they eliminate diverse groups effectively at random and the survivors set the course of subsequent evolution. We concur with Cloud (1948, p. 322) that it is "desirable that workers in several fields of biological science periodically review the principles that they have come to accept and lines of thought that derive from or lead to acceptance of these principles." The issue of periodicity may be just such a case were it would be desirable to review such principles. Maybe it really is time to look to the stars to gain deeper insight into the history of life.

   Part and parcel with the role of astronomical phenomena in general, and cyclical astronomical phenomena in particular, is a debate about the role the physical environment plays in influencing the history of life. This is best viewed in the context of a long standing discourse about the relative roles of biotic versus abiotic forces in evolution: a debate whose focus is beyond the scope of our paper but that goes back at least to Darwin (1859) and probably to von Humboldt (1816), de Candolle (1820), and maybe even Buffon (1749-1804). For specific aspects of the discussion from a paleobiological perspective see Eldredge and Cracraft 1980; Gould and Calloway 1980; Vrba 1980; Ross and Allmon 1990; Vermeij 1993; Benton 1996; Lieberman 2000 and the references therein for a more detailed treatment. Although this debate is beyond the scope of this paper, we would argue that most scientists have moved beyond the view championed by Darwin (1859) and more recently by Dawkins (1976) that much of evolution is caused by competition among organisms. Instead, we would argue that there is ample evidence that at the grand scale history of life the physical environment, construed broadly to subsume both climate and geology, has powerfully structured large-scale patterns of evolution and extinction. For example, there is Valentine and Moores (1970) study showing a relationship between the geometry of the Earth's plates, related to plate tectonic changes, and overall diversity in the fossil record; there are also the more recent studies by Cornette et al. (2002) and Rothman (2001) that showed an association between macroevolutionary origination rate and $CO_2$ levels; finally, there is evidence that animal evolution and extinction have been tracking environmental parameters in largely a random walk fashion throughout much of the Phanerozoic (Cornette and Lieberman 2004).

X.2 Cyclical Astronomical Phenomena Influencing Biology and Evolution: From Weeks to Months to Millions of Years

In the search for evidence that the physical environment plays an important role in structuring evolution, scientists have long recognized the context of our planet in the broader solar system; that is, although not necessarily explicitly recognized, there has been a nascent appreciation for the role astronomical (or astrophysical) phenomena exert upon biology and evolution. For instance, there are diurnal and seasonal cycles (obviously) controlled by the Earth's rotation and revolution and these cause everything from behavioral changes in organisms to migration to selection and thus evolution (Huntley and Webb 1989 provide an outstanding analysis and documentation of these phenomena at several different hierarchical levels). At even longer, decadal time scales there are sunspot cycles that affect the Earth's climate and these may drive certain selection mediated shifts in morphology over similar time scales, for example, the classic studies by the Grant's (e.g., Grant and Grant 2002) on the Galapagos Finches.

Even as the relative time scale grows, astrophysical phenomena remain important influences on extinction and evolution. For example, the signatures of Milankovitch climatic cycles operating on times scales of tens to hundreds of thousands of years are amply evident in the geological record (Hays et al. 1976; Anderson et al. 1984; Olsen 1986; etc.). Moreover, these can influence patterns of stability and change within species lineages, causing stasis, speciation, and extinction (Vrba 1985, 1992; Bennett 1989, 1997). This was lucidly developed by Elisabeth Vrba in her Turnover Pulse Hypothesis.

In a fascinating recent result, Van Dam et al. (2006) even uncovered evidence that cycles of evolution and extinction acted on even longer timescales, roughly 2.5 million years, and these were in turn driven by astronomical phenomena. Vrba (1992) had in fact previously argued for the potential existence of such a cycle, but the evidence was somewhat ambiguous. Certainly more study is needed, but this opens up a potentially interesting window into astronomically governed cyclical phenomena influencing biology and geology that operate on the scale of millions of years. All of these different astronomical phenomena operate at several temporal levels and the larger the time scale the larger the hierarchical level of the genealogical and economic biological entities affected (Eldredge 1999; Lieberman 2000). In conclusion, it seems that from the range of days to months to tens of thousands of years, and possibly even to millions of years, there is enough evidence to posit the stamp of cyclical astronomical phenomena on biology. Thus, it seems not wholly beyond the pale to try to uncover evidence for periodic astronomical phenomena acting on even longer time scales that influence biology. One of the best ways to do this is to try to identify the existence of large scale cycles in the history of life: cycles in diversity or extinction or origination.

X.3 History of Thought on Long-term Cycles of Evolution and Extinction

It turns out that there is a rather long and extensive history of scientists seeking to invoke large-scale (temporal) biological periodicity, and aspects of this history are worth exploring herein. Moreover, the debate about the existence of periodicity in the history of life is itself periodic, and waxes into, and out of, favor over a time scale of roughly every 20-30 years (note, not adduced statistically) with "pulses" of watershed publications occurring in the 1930's, the early 1950's, the late 1970's to the mid-1980's, and the 2000's. We provide a review of the literature on large scale biotic periodicity up through Rohde and Muller's (2005) study, with special emphasis on the earliest studies advocating periodicity; in the next section we then provide a discussion of some of the general criticisms of these studies. Finally, we discuss in greater detail some of the most recent analyses of large scale biotic periodicity (Lieberman and Melott 2007; Melott 2008) and present possible mechanisms that may explain it.

Although the existence of large scale biological cycles do not in and of themselves guarantee the existence of periodic, causal astronomical phenomena (as we shall see, scientists have posited a range of other mechanisms from geology to competition to climate) they may help sway the debate one way or the other, especially when the timescale of the astronomical mechanism matches the inferred evolutionary periodicity.

X.3.1 Amadeus Grabau

The paleontologist A. W. Grabau may have been the first scientist to make a serious attempt to document the existence of this type of periodicity. This was specifically developed in his "pulsation theory," first presented at the International Geological Congress in 1933 and developed in detail in Grabau (1934, 1936). Grabau posited that there were major transgressions and regressions synchronous across all of the Earth's continents related to the swelling and contraction of the seafloor. He further argued that these pulsations would have biological effects, for regression narrows the available area for existence (in the case of marine faunas), which causes a "struggle for existence of ever increasing intensity. This led to the extinction of the weaker and less adaptable, and to the modification of the survivors" (Grabau 1936, p. vii). Eventually, when sea-level rises again, "with renewed expansion, the survivors gave birth to an essentially new fauna which increased in number and variety with the expansion of its habitat" (Grabau 1936, p. vii). The result will be fossil faunas that exist as distinct packages, stacked upon one another in the geological record: "The succeeding transgressive series … contains a fauna which, though it shows certain relationships to that of the preceding retreatal (regressive) series, is on the whole markedly distinct from it, and still more distantly related to the fauna of the transgressive series of the preceding pulsation" (Grabau 1936, p. vii; also see Grabau 1936, p. 4-6). Grabau (1934, 1936) went so far as to comment on the duration of these transgressions/regressions, which he called a "pulse-beat": the "duration of such a pulse-beat is between 20-30 million years" (Grabau 1936, p. 1), an interesting number for several reasons that we shall return to. Further, he was prescient enough to recognize that these pulsations would affect both terrestrial and marine faunas, albeit in different ways.  In particular, transgressions would benefit marine faunas because they expanded the area available to live, yet be to the detriment of terrestrial faunas because they shrank the area available to live.  These ideas were discussed in greater detail, along with additional theories, in Grabau (1940).

X.3.1.1 Relationship between Grabau's pulsation theory and the Turnover Pulse hypothesis

In important respects Grabau's pulsation theory matches aspects of Vrba's (1985, 1992) Turnover Pulse hypothesis.  In particular, each invokes a set of environmental changes that cause deterioration in the environment as a whole, leading to diminishing geographic range size for many taxa; as geographic range sizes decrease eventually extinction and speciation of the ever more narrowly circumscribed taxa occur; finally, as environmental conditions ameliorate, the taxa expand outwards.  There is, however, a difference in the mechanisms precipitating speciation in the pulsation theory and the Turnover Pulse.  In the former, it is competition, following the physical environmental changes, that is the driver: the increasing struggle for existence in an ever narrower geographic region.  By contrast, in Vrba's Turnover Pulse hypothesis it is allopatric speciation that is the driver.  Moreover, Grabau is arguing for significant evolution even as available area expands again, because he claimed that as sea-level rises there is less struggle for existence (in marine faunas), favoring the persistence of less adaptive types (Grabau 1936, p. 4-5) and thus greater diversity.  Again, by contrast, with Turnover Pulse, there should be little if any diversification once climatic conditions improve and species expand their geographic range because the conditions are no longer ripe for allopatric differentiation.  Another difference between their ideas is that the transgressions and regressions Grabau was referring to involved significantly more time then the turnovers Vrba was focusing on: these were related to interglacial/glacial climatic cycles and would involve less than one tenth the time of a pulse/beat.  Still, there is clearly a connection.

X.3.1.2 *Relationship between Grabau's pulsation theory and coordinated stasis*

Indeed, because of this connection, Grabau's pulsation theory also shows similarities to Brett and Baird's (1995) coordinated stasis.  Interestingly, coordinated stasis was developed in the region where Grabau made his early paleontological discoveries: central and western New York State.  Furthermore, the important role that Grabau ascribed to competition in explaining his recovered pattern was also matched in the mechanism Morris et al. (1995) posited explained the pattern of coordinated stasis: ecological locking.

X.3.2 Newell and Simpson on Biological Periodicity

Norman Newell's ideas on periodicity were first presented at the Geological Society of America meeting (Newell 1949) and elaborated on in print in greater detail in Newell (1952). Newell (1949, 1952) identified peaks and troughs of evolution related to both what he called biological factors (read competition), and also physical factors (Newell 1949, p. 1911, 1912; Newell 1952, p. 385). The latter would especially involve the expansion and restriction of shallow epicontinental seaways due to the rise and fall of sea-level. Newell's former advisor, R. C. Moore, also sided with him, arguing that the crinoid fossil record suggested that there was "a spasmodic, pulsatory increase and decrease in the census of species (which) reflect(s) real variations in evolution rate between wide extremes" (Moore 1952, p. 352). Further, Moore (1952) also held that these variations in increase and decrease were likely related to the rise and fall, respectively, of sea level.

However, a subtle to significant difference emerges between Newell's thoughts on this and Grabau's, for Grabau was specifically arguing that it was changes in the physical environment that increased competition. Newell was more equivocal on the cause. Originally, he posited (Newell 1949, p. 1912) that "several groups show similarity of pattern, suggesting common control," perhaps specifically implicating changes in sea-level, but he also argued that although changes in sea-level might cause extinction (through sea-level fall eliminating habitat of marine organisms) they would not lead to an increase in the rate of evolution (Newell 1949, p. 1912). Moreover, whatever initial support Newell (1949) may have had for the control of the physical environment on biological periodicity partly evaporated; he eventually argued that changes in the physical environment will typically not directly affect evolution and further, throughout most of their history, faunas were at adaptational equilibrium (Newell 1952, p. 383, 385).

Although it cannot be decisively shown, the change in Newell's outlook may perhaps have been related to the influence of George Gaylord Simpson, one of Newell's colleagues at the American Museum of Natural History (Niles Eldredge, pers. comm.. 2008), and one of the founders of the Neo-Darwinian synthesis (Eldredge 1985). Simpson, certainly by the late 1940's and early 1950's, completely supported the gradualistic, Neo-Darwinian view, which made him ill disposed to Grabau's ideas on major peaks and troughs of origination and extinction driven by changes in the physical environment. Moreover, his towering intellect and gruff nature made him adept at influencing those surrounding him.

Simpson's viewpoints on this topic were best crystallized in a 1952 publication in the *Journal of Paleontology* derived from a symposium on the "Distribution of Evolutionary Explosions in Geologic Time". Simpson (1952) did not deny the existence of periodicity. For instance, he argued that "(t)here is a well-marked periodicity in vertebrate history, especially as regards successive peaks and valleys of high and low proliferation of new groups" (Simpson 1952, p. 370). However, he also suggested that "it seems impossible even to make a start at realistic correlation of so regular and continuous a process" (Simpson 1952, p. 365), i.e., evolution, with abiotic factors, especially geological changes. Further, he noted that although physical processes likely influenced evolution, they were not decisive (Simpson 1952, p. 369). Instead, biotic factors such as competition were likely to be responsible.

X.3.3 Fischer, Raup and Sepkoski, and Rohde and Muller (2005) on Periodicity

 Fischer and Arthur's (1977) and Fischer's (1982) studies are significant because they were among the first to revive the debate about biological periodicity, and further they suggested an environmental causal connection. In particular, Fischer and Arthur (1977) and Fischer (1982) found support for Grabau's ideas and identified a prominent 32 million year cycle. (Notably, this value is remarkably close to the approximate duration of such cycles inferred by Grabau 1936). Following on the heels of Fischer and Arthur (1977) and Fischer (1982) there are of course the studies by Raup and Sepkoski (1984, 1986) and Sepkoski and Raup (1986) (also see Rampino and Stothers 1984) that identified roughly 26 million year periodicity: a cyclicity that has been attributed to periodic episodes of bolide or comet impact triggered by the motion of a brown dwarf star in our solar system's vicinity (see Raup 1986). (Stanley [1990] presented the interesting perspective that Raup and Sepkoski's retrieved periodicity may have arisen due to the fact that recovery from major extinction events often requires a significant amount of time, either because the hostile conditions that initially caused the extinction continued for some period of time, or the surviving fauna tended to diversify at a low rate: for instance, generalists are more apt to survive an extinction event because of their lower rate of extinction, but tend to also have depressed rates of speciation.) Raup and Sepkoski's (1984, 1986) studies were significant for many reasons, but among the most important was that they were the first to use detailed statistical methods, and a comprehensive database, to try to study this topic.

 The next prominent study supporting periodicity was Rohde and Muller (2005). Their study utilized a revised and updated geological time scale not available to Raup and Sepkoski (1984, 1986) and also incorporated some improved taxonomic data in the "Sepkoski dataset" (Sepkoski 2002). They failed to find evidence for a cycle at 26 (or 32) million years but they did find strong support for a cycle in biotic diversity operating every 62 million years. Rohde and Muller (2005) largely left the issue of causal mechanisms for such a long duration cycle open, though Rohde (2006) subsequently suggested the causes were related to geological phenomena.

X.4 The Problem of Uniformitarianism and the Completeness of the Fossil Record

It is worth noting that just as there has been a long history of identifying evidence for periodicity in the history of life there has been a long history of challenging that very evidence. This typically happens anytime a novel theory or idea is developed and therefore is not that surprising. Some of the motivation for scientists' opposition to periodicity seems to have been partly a matter of metaphysics and philosophy. A clear example of this comes from the symposium on the "Distribution of Evolutionary Explosions in Geologic Time" in the *Journal of Paleontology* that was already mentioned. Although Newell (1952) and Simpson (1952) there endorsed aspects of periodicity (although they did not endorse Grabau's causal mechanism) on the whole the general tenor of the symposium was a rejection of periodicity, and Cooper and Williams' (1952) paper is typical in this regard. However, some of the authors went so far to reject the very notion that there might be times of explosive evolution or extinction necessitated by periodicity (e.g., Camp 1952). Partly this rejection was done on the grounds that variations in rates of evolution and extinction, and therefore periodicity, violated the basic tenets of aspects of uniformitarianism (see Gould 1965). This was related to the basic Neo-Darwinian presumption, prevalent at that time, that evolutionary change (and extinction) must always be slow and gradual. Consider that Simpson (1952), one of the architects of the Neo-Darwinian synthesis, and by then a committed gradualist (Eldredge 1985), argued for the continuous rather than pulsed nature of evolutionary origins and also downplayed the existence of mass extinctions in the fossil record.

In some respects, we see parallels between how Grabau's works were treated by certain paleontologists in the 1950's and how Goldschmidt's works were treated by certain evolutionary biologists (including the paleontologist George Simpson) in the 1940's. Goldschmidt was the bête noire of the Neo-Darwinian synthesis because his idea of "hopeful monsters" (Goldschmidt 1940) was at odds with certain population genetic principles. Now, certainly scientists opposed aspects of Goldschmidt's work for legitimate reasons, and their were several problems with his reasoning, but what really got many scientists' goad was that to Goldschmidt evolution could not be explained by the simple extrapolation of microevolution to macroevolution: Goldschmidt's view of evolution was not "uniformitarian" enough (Gould in Goldschmidt 1982). Similarly, at times Grabau and his "pulsation theory" seems to have become a punching bag because he was espousing ideas that invoked significant variations in rates of evolution and extinction due to variations in the physical environment. Yes, Grabau's pulsation theory may have been outrunning the available evidence (to paraphrase Henbest 1952, p. 318) but partly what caused his theory to be substantially criticized was the fact that some of his ideas were anathema to Neo-Darwinian uniformitarianism.

Other criticisms of periodicity were data and analysis driven, for example, the criticism of Raup and Sepkoski's analyses by Patterson and Smith (1987). Patterson and Smith (1987) legitimately challenged aspects of Raup and Sepkoski's dataset for they discovered that Raup and Sepkoski's extensive (though inadvertent) inclusion of paraphyletic taxa may have clouded their results. Other critiques have surfaced based on the types of methods used to adduce periodicity (see discussion in Sepkoski 1989 and also in Lieberman and Melott 2007, the latter citation discussed more fully below).

Another frequently offered reason for rejecting the existence of periodicity was that there were too many problems with the fossil record (it is too biased or too poor) or that the paleontological data themselves (that is, data relating to taxonomic diversity, and data relating to biostratigraphic and chronostratigraphic correlations) were too poor such that evidence for periodicity could not be trusted (see, e.g., Henbest 1952 and also Camp 1952 and Cooper and Williams 1952 in reference to Grabau's 1936 conclusions and Stanley 1990, in reference to Raup and Sepkoski's 1984, 1986 conclusions). Indeed, from the references here it is clear that this type of criticism of periodicity, that the fossil record (or paleontological data) is (are) compromised, has repeatedly resurfaced. Of course the fact that the fossil record is not perfect has been raised since at least the time of Darwin (1859) and it surely represents a legitimate card to play to explain our inability to see something in the fossil record; it may even represent a legitimate card to play when explaining why we do see something in the record (i.e., perhaps periodicity could arise of an artifact of the geological record). From Raup and Sepkoski's (1984) study up through Rohde and Muller's (2005) study the data utilized were various permutations of the so called "Sepkoski" dataset (most recently collected in Sepkoski 2002). Discussion of the validity of the Sepkoski dataset is beyond the scope of this paper. There have, however, been several studies that have criticized the dataset, and questioned aspects of its validity. Further, there have been a variety of studies aimed at addressing whether or not the Sepkoski dataset reflects true variations in diversity through time or rather variations in the processes of the geological record that preserve that diversity (e.g., Adrain and Westrop 2000; Alroy et al. 2001; Peters 2005; Foote 2006; see also discussion in Bambach 2006). Based on these issues, several new taxonomic databases, have been developed, the most important being the Paleobiology Database (PBDB), which we will discuss more fully later in the paper. We would argue that uncovering *either* evidence for periodicity in true diversity or in preserved diversity relating to artifacts of preservation of the geological record would be interesting, albeit for different reasons. The former would imply that there was some periodic mechanism controlling the diversity of life on our planet; the latter, that there was some periodic mechanism controlling the geological processes governing the fossil record.

The issue with the poor quality of paleontological data in general is an interesting one. (Partly it devolves to an issue of what evolutionary principles the fossil record can be used to study, and that topic seems beyond the scope of this contribution; however, we certainly would argue that the fossil record is a valuable repository, containing much information useful for evolutionary studies and further are our one true chronicle of the history of life.) Raup and Sepkoski (1986, p. 833) did argue that if the data were poor the result should degrade toward randomness. In particular, they claimed that uncertainty in databases of paleontological diversity likely would not lead to periodicity, and this seemed reasonable, yet by the same token the result of Patterson and Smith (1987) already described provided a cautionary tale.

X.5 Re-Analyzing Evidence for 62 and 26 million Year Periodicity

X.5.1 Summary of Lieberman and Melott's (2007) analysis

Partly motivated by our desire to assess the resiliency of Rohde and Muller's (2005) evidence for periodicity (and also earlier analyses of periodicity) we (Lieberman and Melott 2007) decided to consider their study in greater detail. In particular, we were interested in determining if their results changed when additional and supplementary statistical tests were applied. First, the data from the Sepkoski (2002) dataset (the dataset used by Rohde and Muller 2005) were detrended; this is essential in any analysis looking for periodicity when there is an overlying trend (in this case diversity is increasing through time); details on detrending are provided in Lieberman and Melott (2007). (Note, Rohde and Muller 2005 and Cornette 2007 also employed detrending.) Rohde and Muller (2005) used Fourier Spectral Analysis (Brigham 1988) to uncover evidence for periodicity in fossil biodiversity. Lieberman and Melott (2007), by contrast, used the Lomb-Scargle Fourier Transform (Scargle 1982; Laguna et al. 1998) which is more effective at analyzing time series data where the data points are not evenly spaced. (In the case of fossil diversity data, the samples are not evenly spaced in time.) (More details on the analysis are provided in Lieberman and Melott 2007.) Our results built on the analysis of Cornette (2007) who used the equivalent Gauss-Vanìcek method to re-analyze Rohde and Muller (2005). It is worth mentioning that any (non pathological) function can be decomposed into a sum of sine waves. The interesting question is whether any stand out above the others, with much higher amplitude. Fourier Analysis is a highly standardized way to approach this, but it requires evenly spaced data. The irregularly spaced data of the fossil record can be used, but interpolation is needed. This can introduce artifacts if not understood by the practitioner. Lomb-Scargle/Gauss-Vanìcek is based on doing a least-squares fit to sinusoids, and the data do not have to be evenly spaced so no interpolation is necessary. The safest thing to do is to use both methods, which was done by Lieberman and Melott (2007). It turns out Rohde and Muller (Muller, personal communication 2008) did it also but did not report the check.

Rohde and Muller (2005) only analyzed biodiversity. Lieberman and Melott (2007) also analyzed Phanerozoic biodiversity since 519Ma, but in addition considered fractional biodiversity (which considers the relative amount of a diversity change at a given time interval and thus may be more biologically significant). Further, we investigated whether various individual mass extinction events were contributing disproportionately to the evidence for cyclicity. Raup and Sepkoski (1986) and Cornette and Lieberman (2004) indicated this was a possibility. We also investigated whether cyclicity was primarily implicit in the pre-150Ma or post-150Ma parts of the data as this appears to mark an important transitional period in the history of life (Bambach 2006). Finally, we investigated origination and extinction (and various metrics of these); Rohde and Muller (2005) also considered these, though in a somewhat different manner (see Lieberman and Melott 2007 for further discussion).

X.5.1.1 Summary of Lieberman and Melott's (2007) results

Even using the different analytical method, Lieberman and Melott (2007) continued to find strong evidence for periodicity in biodiversity (significant at the .01 level) and also in fractional biodiversity at roughly $62 \pm 3$ Myr (Fig. 1); the latter is significant at the .001 level, matching Rohde and Muller's (2005) identified peak but at a higher level of significance; there is also a peak at roughly $31 \pm 1$ Myr: it closely matches Fischer and Arthur's (1977) and Fischer's (1982) peak (and thus also near Grabau's more informally defined 1936 pulse), although it is only significant at the .1 level and thus not treated as statistically significant; this peak, however, may merit further study. Notably, even removing the Ordovician/Silurian, Permo/Triassic, and Cretaceous/Tertiary mass extinctions from the biodiversity data did not eliminate the strong 62 Myr peak.

Periodicity in biodiversity (and fractional biodiversity) seems to be primarily confined to the time interval 519-150Ma; post 150Ma the data show no evidence for periodicity (results not shown), but prior to 150Ma there is a cycle at roughly $61 \pm 3$ Myr very close to Rohde and Muller's (2005) 62 Myr cycle, significant at the .001 level, and also a cycle at roughly $32 \pm 1$ Myr very close to Fischer and Arthur's (1977) and Fischer's (1982) cycle (and thus also in the window of Grabau's 1936 cycle) significant at the .05 level (Fig. 2).

The results from analysis of certain metrics of origination reveal significant (at the .01 level) cycles at roughly $60 \pm 3$ Myr and $24 \pm .5$ Myr: the former close to the Rohde and Muller (2005) cycle; the latter close to Raup and Sepksoki's (1984, 1986) cycle (and thus also in the window of Grabau's 1936 cycle) (although in origination, not extinction) (Fig. 3). The results from analysis of extinction metrics reveal a cycle at roughly $27 \pm 1$ Myr significant at the .02 level (results not shown): effectively indistinguishable from Raup and Sepkoski's (1984, 1986) cycle (and again thus also in the window of Grabau's 1936 cycle) and covering a greater interval of time (Raup and Sepkoski primarily concentrated on the Mesozoic and Cenozoic fossil record). However, relevant for the analysis of extinction, Stigler and Wagner (1987) presented the interesting result that there was a peak in stratigraphic interval length at roughly 26 Myr which could have caused Raup and Sepkoski (1984, 1986) to artifactually retrieve a peak at that time interval. Lieberman and Melott's (2007) reanalysis of the data confirms a peak at roughly $27 \pm 1$ Myr in stratigraphic interval length significant at the .001 level (Fig. 4). This could imply that peak is an artifact, or that there is a process operating roughly every 27 Myr that causes extinctions to occur and leaves the fossil record broken up into intervals that stratigraphers subsequently identify (see more extensive discussion in Sepkoski 1989 and Lieberman and Melott 2007). However, there is no peak in stratigraphic interval length at or near 62 Myr suggesting that peak, retrieved in the other analyses, is not an artifact (Fig. 4).

In conclusion, it would appear that based on the analysis of the Sepkoski (2002) dataset there is strong evidence for Rohde and Muller's (2005) identified periodicity in biodiversity at 62 Myr; there is more equivocal evidence for periodicity in biodiversity at roughly 32 Myr (vindicating Grabau 1936 and Fischer and Arthur 1977 and Fischer 1982). There is also evidence for periodicity in origination at roughly 60 Myr (supporting Rohde and Muller's 2005 identified peak) and also roughly 24 Myr (supporting Grabau 1936 and indirectly Raup and Sepkoski 1984, 1986, because they emphasized extinction while Grabau emphasized pulses of origination and extinction). Furthermore, there is evidence for periodic extinctions at roughly 26 Myr (again vindicating Grabau 1936, and also Raup and Sepkoski 1984, 1986, but with the important caveat about stratigraphic interval lengths mentioned). Finally, it is worth stating the primary evidence for periodicity comes from the Paleozoic and early to mid Mesozoic fossil record. This could signify the result may be artifactual, if these data are of poorer quality than more recent data, it could signify that there are some biological differences between post and pre- 150 Ma organisms, or it could be that the mechanism that once caused periodicity suddenly disappeared post 150 Ma. In any event, it would seem that there is at least some evidence supporting the notion that one or more large scale cycles, operating on time frames of tens of millions of years, have influenced the history of life.

*X.5.2* Summary of Melott's (2008) results

Because the validity of the Sepkoski dataset has been perennially criticized, one of us (ALM) performed an additional test to look for evidence of long term periodicity (Melott 2008); this analysis used the PBDB, which attempts to correct the fossil record of diversity, and specifically improve on the Sepkoski dataset, by using various sample standardization techniques, in addition to other improvements.

Melott performed an analysis similar to that of Lieberman and Melott (2007) but using the PBDB data provided by John Alroy. He found that both the FFT and Lomb-Scargle analyses showed the existence of a 63 Myr cycle in the PBDB data. The fact that PBDB contains strong corrections for sampling rate suggests that sampling rate is not a causal factor in the existence of this period. Melott (2008) also shows a *cross-spectrum*, which is a technique for finding similarities in the periodicities of two different time series; in this case it was the data used by Rohde and Muller (2005) and the PBDB data. The cross-spectrum showed a strong peak at 62 Myr, and indicated that not only the period but the timing of the peaks and valleys of the cycle in the two different databases only disagreed by an average of 1.6 Myr. Again, this provides strong support for the notion that there is a strong signature of long term periodicity present in the history of life.

X.6 Possible Causal Mechanisms for 62 Myr Cyclicity

Rohde and Muller (2005) discussed, without endorsing, a wide variety of possible mechanisms that might cause a fairly regular swing in biodiversity over roughly the last 500 Myr. They basically looked at two types of mechanism: (1) those than can easily impact biodiversity, but for which there is no reason to expect such periodicity; (2) periodic phenomena which have no obvious connection to biodiversity.

Astronomical processes are the logical place to look for regular periodic processes over long timescales. Most large-scale motions in the universe are driven by gravity, a weak force acting over large distances, which naturally gives rise to long time periods. Bound systems with insignificant friction (space being mostly empty) give processes whose periods are very stable over long times. This leads to the question of what kind of system might be periodic over 62 Myr.

Processes to do with the Earth-Moon system of course take months, and inner Solar-system processes take years. There are distant, gravitationally bound objects called the Oort Cloud, which are almost certainly the origin of long-period comets. The orbits of these objects may take up to a million years. They could be perturbed, and fall into the inner Solar System, if the Sun had a dim stellar companion with a very long-period orbit (Whitmire and Jackson 1984; Davis et al. 1984). However, any object with such a long timescale orbital period would have a larger orbital radius, and be only weakly bound to the Sun by gravity. Passing stars and molecular clouds would likely change the period of such an object, possibly even dislodging it from it orbit (Hills 1984). Therefore, these do not appear to be viable candidates.

Motions of entire galaxies across space take billions of years. For example, our Local Group contains satellite galaxies, but most have orbital periods which are quite long and also not terribly regular due to the complicated structure of the group. Thus, these motions are not good candidates, because the timescales are too long and too irregular.

This leaves the motions of stars within the Galaxy, and here we find a good candidate (Medvedev and Melott 2007). The galaxy is a thin disc, with the Sun and its planets located about two-thirds of the way out from the center. The Sun orbits the center of the Galaxy with a time period of about 200 Myr. As it does so, it will move in and out of spiral arms, which are areas of temporary concentration of stars and gas. Such regions have enhanced star formation, and that means they have many supernovae, which come from large, hot, short-lived stars. There is enhanced danger to the biosphere during such crossings; alas the spiral structure is not very well-known, but not likely to be terribly regular.

There is, however, another motion which fits closely a 62 Myr periodicity: as the Sun moves around the Galaxy, it also executes a wobbling up and down motion, rather like a weight bobbing on a spring (Medvedev and Melott 2007). (In fact, due to a peculiarity of the distribution of mass in the Galaxy, this motion even behaves mathematically in a way not far from that of a weight on a spring). Gies and Helsel (2005) have provided a solution for the Sun's motion in the past given our knowledge of its present position, velocity, and the gravitational field of the Galaxy. Interestingly, the period of the vertical motion is about 63 Myr. This suggests an interesting possible coincidence. But of course, there has to be some kind of effect. The most natural thing is something that happens when we pass through middle of the disc, where the mass is more concentrated. There are two problems with this. First, given the approximately 200 light-year amplitude of our "vertical" motion, the mass is not all that much more concentrated at the center than at the extremes. Secondly, something that happens then would happen *twice* per period, or every 32 Myr. It is possible that the various weaker periodicities in the vicinity of 30 Myr which we have mentioned earlier have to do with passage of the Sun through the Galactic midplane on this timescale (e.g. Matese 1996), resulting in an increase in the rate of comet impacts on the Earth. But this could not explain the stronger 62 Myr signal.

In order for this to be responsible for 62 Myr periodicity, something has to happen once per period, which would mean on one side of the galaxy, or when passing through the disc going in one direction. The galaxy itself is reasonably symmetrical, so this doesn't seem to make much sense. However, the external environment isn't symmetrical. The local Universe has a lot more mass on one side of our galactic disc than the other. In particular, the Virgo Cluster of galaxies is the only large mass concentration in our vicinity, and its gravitational attraction has given our galaxy a speed of about 200 km/s, falling toward it (Medvedev and Melott 2007). This cluster is located only about 16° off the Galactic north axis. Furthermore, an examination of the detrended biodiversity information plotted against the Gies and Helsel (2005) Solar motion (Fig. 5) shows that biodiversity declines tend to coincide with excursions to galactic north—toward the Virgo Cluster; the cross correlation is significant at $p = 2 \times 10^{-7}$ (see Medvedev and Melott 2007). That is, not only do the periods of the two cycles agree, but the timings of the peaks agree very strongly as well.

Medvedev and Melott (2007) noticed this coincidence and proposed a mechanism for biodiversity fluctuation. They suggested that the 200 km/s infall toward the Virgo Cluster will produce a shock wave, and push it toward the galaxy on the north side. This is a shock wave in the plasma (hot ionized gas) which fills space, not too different from the shock that precedes a supersonic jet or the shock wave that the Solar System produces as it moves through the galaxy at a few km/s. A shock in plasma is well-known to produce cosmic rays, which are protons and other charged elementary particles traveling at very high speeds. We are normally exposed to some cosmic rays, which account for a large fraction of the radiation dose we get.

Medvedev and Melott (2007) proposed that the cosmic ray flux at the Earth would increase greatly when we move to galactic north. Thus the times of low biodiversity would correspond to times when the Earth is exposed to a high level of cosmic rays. The increased flux would come because we have moved closer to the source, the shock front, but even more so because there is less magnetic shielding between the Earth and the source.

Detailed computation of the spectrum and intensity of the cosmic rays and their resulting effects has not been done. However, some generalizations are possible. (1) Cosmic rays are a known mutagen, and as such can contribute to genetic change, including cancer. A large increase in the rate of such events might have an effect to lower biodiversity. (2) Cosmic rays ionize the atmosphere. Such ionization can change its chemistry, increasing the concentration of oxides of nitrogen. These in turn catalyze the destruction of ozone ($O_3$). Ozone is the primary shield that keeps destructive Solar UVB radiation away from the Earth's surface. UVB can cause cell damage and break DNA molecules. (3) There is increasing evidence that the same atmospheric ionization can provide nucleation sites for cloud formation. It has been hypothesized that increased cloud formation will change the albedo of the Earth, lowering temperatures. All these possible mechanisms are plausible, but they all need a lot more theoretical work and comparison with data, where possible, before being declared good candidates. Unfortunately, at this time there are no known long-term isotopic changes that offer hope of a direct measurement of the cosmic-ray variability.

X.7 Conclusions

Our sense is that there is tantalizing but diffuse evidence for the possible existence of long term cycles of biodiversity, possibly related to astronomical phenomena. This evidence has been accumulating over the course of many decades and is part of a larger debate about the role the physical environment plays in causing evolution and also about uniformitarianism. We acknowledge that the issue clearly has not been demonstrably proven by any sense of the imagination, and even the most recent round of studies by Rohde and Muller (2005), Cornette (2007), Lieberman and Melott (2007), and Melott (2008) is unlikely to tip the balance, though it may be useful that a possible mechanism to explain 62 Myr periodicity has now been uncovered. The truth is that this issue may never be definitively demonstrated: our focus should be more on testing hypotheses in any case. However, even though paleontology tends not to be a predictive science we will make one prediction that the future may, or may not, bear out: if the most recent studies on periodicity do not cinch the matter, then we are in for another round of studies focusing on long term biological periodicity roughly sometime in the mid-2020's. In any event, the existence of such cycles would imply that the physical environment, in particular, the environment external to our planetary biosphere, has had a profound and continual influence on macroevolutionary patterns. This may speak to the notion that contingency (*sensu* Gould, 1989) rules the history of life, yet there is some repetitive nature to those contingent events.

X.8 Acknowledgments

We thank NSF DEB-0716162 (to BSL) for support of this research and N. Eldredge and M. Medvedev for discussions.

**Referencees**

**Figure Captions**

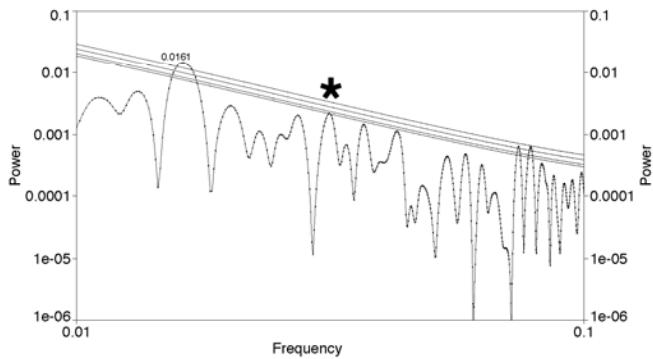

Figure 1. Analysis of detrended fractional fossil biodiversity fluctuations from Lieberman and Melott (2007). Lines denote 0.1, 0.05, 0.01, and 0.001 levels of significance. Frequencies are given per Myr; there is a peak at a frequency of approximately 0.0161/Myr which is equivalent to 62.1 ± 3.1 Myr; this peak is significant at better than the .001 level. There is also a peak marked by the "*" at roughly 31 ± 1 Myr, although it is only significant at the .1 level. All other significant peaks occur at less than 15 Myr and thus are not relevant because they are below the so called Nyquist frequency (see Lieberman and Melott 2007 for further discussion).

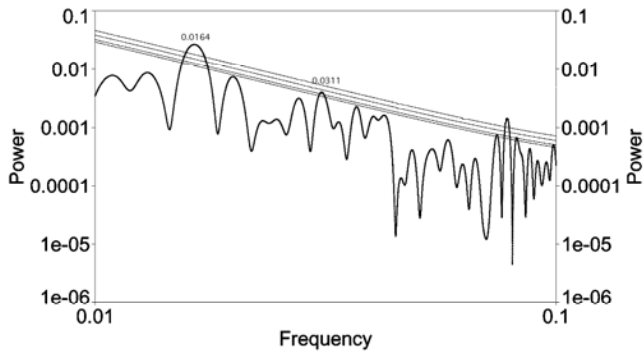

Figure 2. Analysis of detrended fractional fossil biodiversity fluctuations from 150-519 Ma from Lieberman and Melott (2007). Lines denote 0.1, 0.05, 0.01, and 0.001 levels of significance. Frequencies are given per Myr; there is a peak at a frequency of approximately 0.0164/Myr which is equivalent to 61.0 ± 3.2 Myr; this peak is significant at better than the .001 level. There is also a peak at a frequency of roughly 0.0311Myr, equivalent to 32.2 ± 1.1 Myr, although it is only significant at the .1 level. All other significant peaks occur at less than 15 Myr and thus are not relevant because they are below the so called Nyquist frequency (see Lieberman and Melott 2007 for further discussion).

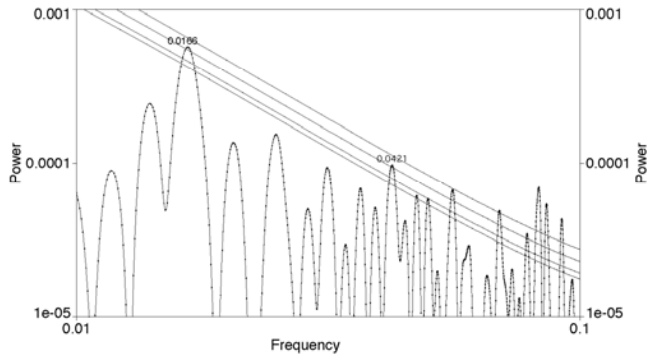

Figure 3. Analysis of fluctuations in fractional origination intensity from Lieberman and Melott (2007). Lines denote 0.1, 0.05, 0.01, and 0.001 levels of significance. Frequencies are given per Myr. There are two peaks significant at the .01 level: one at a frequency of approximately 0.0166/Myr which is equivalent to 60.1 ± 3.1 Myr; one at a frequency of roughly 0.0421Myr, equivalent to 23.7 ± 0.5 Myr. All other significant peaks occur at less than 15 Myr and thus are not relevant because they are below the so called Nyquist frequency (see Lieberman and Melott 2007 for further discussion).

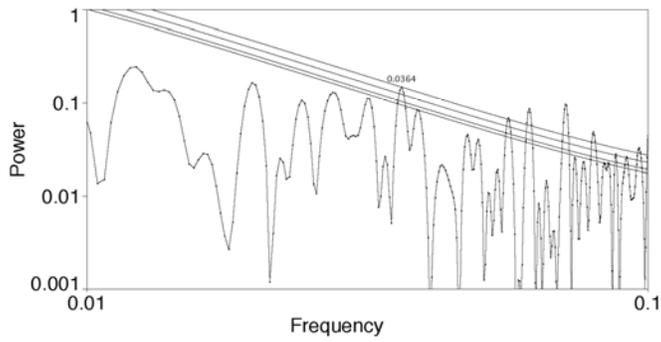

Figure 4. The power spectrum of stratigraphic interval lengths from Lieberman and Melott (2007). There is a peak significant at the 0.001 level at a frequency of roughly 0.0364/Myr, equivalent to 27.5 ± 0.6 Myr; note there are no peaks at or around 62 Myr.

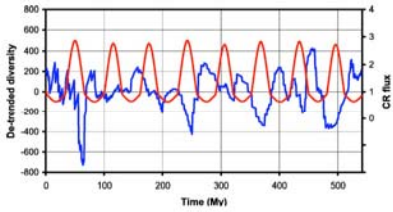

Figure 5. Detrended diversity variation (*blue curve, left scale*) as a function of time. Also, normalized cosmic-ray flux calculated from Medvedev and Melott (2007), based on their model of solar system position in galactic arm (*red* curve, right scale). The maxima in the cosmic-ray flux coincide with minima of the diversity cycle. Note also that the onset of diversity decline coincides with rapid increase of the flux. The cross correlation is significant at $p = 2 \times 10^{-7}$ (see Medvedev and Melott 2007).